\documentclass[12pt,preprint]{aastex}
\slugcomment{To appear in the Astronomical Journal}
\shorttitle{Proper motion measurements of twenty-eight pulsars}
\begin{document}
\title{Proper-Motion Measurements with the VLA. II. Observations of
  Twenty-eight Pulsars}

\author{W. F. Brisken,\altaffilmark{1} A. S.
  Fruchter,\altaffilmark{2} W. M. Goss,\altaffilmark{1} R. S.
  Herrnstein,\altaffilmark{3} and S. E. Thorsett\altaffilmark{4}}
\altaffiltext{1}{National Radio Astronomy Observatory, P.O. Box 0,
  Socorro, New Mexico 87801 wbrisken@nrao.edu; mgoss@nrao.edu}
\altaffiltext{2}{Space Telescope Science Institute, 3700 San Martin
  Dr., Baltimore, MD 21218 fruchter@stsci.edu}
\altaffiltext{3}{Columbia Astrophysics Laboratory, Mail Code 5247,
500 West 120th St. New York, NY 10027}
\altaffiltext{4}{Department of Astronomy and Astrophysics, University
  of California, 1156 High St., Santa Cruz, CA 95064 thorsett@ucolick.org}

\clearpage

\begin{abstract}
  Using the Very Large Array, we have measured the proper motions
  of twenty-eight radio pulsars. On average, the pulsars studied are
  fainter and more distant than those studied in earlier work, reducing
  the selection biases inherent in surveys restricted to the Solar
  neighborhood. The typical measurement precision achieved is a few
  milliarcseconds per year, corresponding to a few tens of kilometers
  per second for a pulsar a kiloparsec away.  While our results compare
  well with higher-precision measurements done using very-long baseline
  interferometry, we find that several earlier proper motion surveys
  appear to have reported overly optimistic measurement uncertainties,
  most likely because of a failure to fully account for ionospheric
  effects. We discuss difficulties inherent in estimating pulsar velocities
  from proper motions given poorly constrained pulsar distances.
  Our observations favor a
  distribution with 20\% of pulsars in a low velocity component 
  ($\sigma_\mathrm{1D} = 99$~km~s$^{-1}$) and 80\% in
  a high velocity component ($\sigma_\mathrm{1D} = 294$~km~s$^{-1}$).  
  Furthermore, our sample is consistent with a scale height of pulsar 
birthplaces comparable to the scale height of the massive stars that 
are their presumed progenitors.  No evidence is found in our data for a 
significant population of young pulsars born far from the plane. We 
find that estimates of pulsar ages based on kinematics agree well with 
the canonical spin-down age estimate, but agreement is improved if 
braking indexes are drawn from a Gaussian distribution centered at $n$=3 
with width 0.8.

\end{abstract} 

\clearpage

\section{Introduction}

Radio pulsars are a high velocity stellar population, with typical
speeds of hundreds of kilometers per second, induced by asymmetries in
their birth supernova events or through breakup of progenitor binary
systems.  Although most pulsars are probably born very near the plane
of the Galaxy, they can move to a kiloparsec or more above the plane
in a few million years. Most proper motion studies have been limited
to the bright, nearby pulsars in the Solar neighborhood, and therefore
miss a significant fraction of the highest velocity stars. This bias
must be considered when attempting to estimate the birth velocity
distribution from the observations. The initial velocity distribution
contains valuable information about supernova symmetry and the
progenitor binary properties and is an essential input parameter to
binary population synthesis studies and for calculations of the
neutron star retention rate in globular clusters.  Considerable effort
has therefore been expended in trying to understand the effects of
survey incompleteness on the observed pulsar velocity distribution.

An alternative approach is to reduce the bias by expanding the volume
of the Galaxy in which pulsar proper motions are measured. In 1992, we
set out to do that with a long-term study of a sample of relatively
distant pulsars, using the NRAO\footnote{The National Radio Astronomy
Observatory is a facility of the National Science Foundation operated
under cooperative agreement by Associated Universities, Inc.}  Very
Large Array (VLA) telescope facility.  In McGary et al. (2001, Paper~I), we
briefly described the motivation for this work and our target
selection criteria, and we described our observational method, which
we review in \S\ref{sec:obs}.  In this report, we present proper
motion measurements for twenty-eight pulsars (\S\ref{sec:results}).
In a few cases, previous measurements with lower precision are
available from other surveys, allowing us to test the reliability of
these surveys.

Velocity estimates require combining our results with distance
estimates. In \S\ref{sec:dist}, we discuss the importance of distance
uncertainties in estimating pulsar velocities and we estimate the
distances of the pulsars in our sample. Finally, in \S\ref{sec:disc}
we present some discussion of the implications of this work.  We
calculate the relatively bias free velocity distribution using a 
sample of 36 pulsars with reliable proper motion measurements.
Additionally for the young pulsars of this sample we compute the 
probability distribution function (PDF) for
kinetic age and hence compute the fraction of young pulsars that appear to
be falling toward the Galactic plane.

\section{Observations \label{sec:obs}}

The observations and data analysis techniques have been discussed in
detail in Paper~I, and are briefly reviewed here.

We observed 28 pulsars during observing campaigns at the VLA centered
at epochs 1992.96, 1994.23, 1995.52, 1998.32, and 1999.47. 
Most observations were made in the A-array, with a typical beam size of
1.3~arcsec.  A small number
of observations were made during the reconfiguration to 
B-array in 1998, with beam sizes up to 4~arcsec.
Only the two circular polarization products,
RR and LL, were simultaneously correlated; the linear polarization products
were not retained as that would have reduced the available bandwidth without
increasing the sensitivity.
A 25~MHz wide band at 1452~MHz was
divided into 15 frequency channels to reduce bandwidth smearing over a
wide field, which allowed identification of multiple in-beam
astrometric reference sources for each pulsar.  Typical image noise 
levels ranged between 0.1 and 0.2~mJy beam$^{-1}$.  In many cases, the VLA
correlator was ``gated'' for one polarization, to accumulate data only
during the on-pulse portion of the pulsar period.  By eliminating the
accumulation of noise during the off-pulse portion, the signal-to-noise
ratio was improved by up to a factor of five.  This, together with the
high sensitivity of the VLA, allowed observation of fainter---and more
distant---sources than in previous studies.  The second polarization
was ungated to maximize sensitivity to the reference sources. We
performed tests to verify that the astrometry was not affected by the
gating, comparing the position measurements from the two polarizations.
Agreement to 5~mas or better was achieved for the brightest reference
sources; statistical uncertainties dominated for the weaker sources. 
For the brighter pulsars the gate was not used and the two
polarizations were summed. A summary of the pulsar observations can be
found in Table~\ref{tab:obs}.  

Earlier VLA observations made at $\lambda \sim 21$~cm
were also available for four of our target pulsars.  We
reanalyzed the archival raw data.  Most of these data were
observed for previous position or proper motion surveys 
(in particular, Fomalont et~al. 1992 \& 1997).
These observations proved valuable as they roughly doubled the time span of
some of
the position measurements.  Including these data was not entirely 
straight forward, however.
Because these earlier observations were
made with a 50~MHz bandwith divided into fifteen channels, the effect of
bandwidth smearing is about twice as great.  The positional
uncertainties of reference sources far from the field center were increased
in the radial direction to properly account for this.
The observation frequency for these additional observations was 1385~MHz, 
rather than 1452~MHz, resulting in a 5\% larger beam, 10\% larger
ionospheric effects, and possibly slightly different structure in the
reference sources.  None of these effects is great enough to affect the
results.  In the end, addition of the archival data did not appreciably 
change the proper motion results (typical improvement was 20\%).  
This is mainly due to less
time on source than for the new observations and to the lack of gating.  
Although an earlier observation of B1237+25 was made, it was not included
in our analysis.  These archival observations 
did provide additional confidence that the proper motions values
reported here are correct.
Additionally, B2106+44 was observed as a test target with the VLA and
the Pie Town VLBA antenna in 2001 and these data were included in our
analysis.  

Flux density and bandpass calibration were obtained from observations
of 3C48, 3C147, or 3C286.  Observations of each pulsar were alternated
with observations of a nearby ($<20^{\circ}$) phase calibrator. Phase
and amplitude solutions were interpolated between calibrator
observations and applied as complex gain corrections to the pulsar
visibilities. Self calibration was performed in all cases.  Self
calibration is an iterative process that leads to self-consistent
calibration resulting in a maximized signal-to-noise ratio.
For gated measurements, the
self-calibration solutions from one polarization were copied to the
other polarization. Uncertainties in the phase calibrator position and
differential ionospheric and tropospheric phases limit the absolute
pulsar position measurements to about 0.1~arcseconds.

Analysis was done in the AIPS software package.  The task UVFIX was
run to recover the correct $(u, v, w)$ values and timestamps for the
source.  This is needed for high-precision astrometry since the VLA
correlator does not include special relativistic corrections for
diurnal and annual abberation and the timestamps reported are for the
start of integration, not the midpoint (see Fomalont et~al. 1992 for
discussion on this topic).  Images were made using the
CLEAN algorithm.  Self-calibration was used to improve estimates of
the phase and amplitude errors.  Self-calibration can reduce the accuracy
of absolute position measurement, however the relative astrometry,
used here to measure the motions of the pulsars, is 
improved due to the increase in sensitivity.
JMFIT was used to fit a Gaussian brightness distribution
to the pulsar and the compact reference sources producing position
measurements and their uncertainties.  An error-ellipse for each position
measurement was computed in order to correctly account for covariances
in the $x$ and $y$ position uncertainties.  See Paper I for more details
on our use of UVFIX and JMFIT.  

\section{Proper Motions and Uncertainties\label{sec:results}} 

Pulsar positions and proper motions were determined through a 
$\chi^2$ minimization that simultaneously fit for the 
positions of each of the
reference source and the pulsar, the proper motion of the pulsar,
and a small residual systematic coordinate deformations relating each epoch's
coordinate system to that of the first epoch.  
The magnitude of these coordinate deformations were strongly dependent on
the source declination, suggesting
differential refraction through the ionosphere and
troposphere (see Paper~I) as a cause.  The transformation 
was parameterized as a six parameter linear transformation.  
In the cases where the fit residuals were high, the reference sources were
investigated.  Those that showed proper motion significant at the $2 \sigma$ 
level or very high fit residuals were removed.  In all cases, removal of between zero and three reference
sources from the fit produced residuals that were consistent with the
measurement uncertainties.  These sources likely experienced structure changes,
such as ejection of material into a jet or relative brightness changes of
components, that are unresolved with the VLA but none-the-less produce 
astrometic shifts.
For pulsars with four or more epochs, uncertainties in
the proper motions could also be estimated using bootstrap Monte-Carlo
techniques \citep{pftv86}. Where this was possible, the agreement
between the methods was excellent, and we believe that the
uncertainties derived from the least squares fitting are accurate,
even for the less-sampled sources.  We believe that our final errors
are dominated by the Gaussian measurement uncertainties and that the
final error bars are very nearly the Gaussian $1\sigma$ widths. The
astrometric results are summarized in Table~\ref{tab:res}.
See Paper~I for additional details on the fitting and Monte-Carlo methods.
It should be noted that no strong covariances between the proper motion
and other fit parameters were observed.

For all but three of the pulsars observed, these new results were
either the first or the most accurate proper motion measurements.  Two
of the exceptions, B0919+06 \citep{ccl+01} and B1237+25 \citep{bbgt02}
were observed using the Very Long Baseline Array (VLBA) using proven
techniques.  Single-epoch VLBA position measurements of these pulsars
were accurate to about 0.2 mas.  The third, B1534+12, is a stable
millisecond pulsar with a very accurate timing proper motion
\citep{sac+98}.  Except for a single $2\sigma$ deviation, in the 
measurement of $\mu_\alpha$ for B1237+25, our new proper motion 
measurements were all within one standard deviation of these more 
accurate results.  These results are consistent with Gaussian statistics.

The new proper motions are compared against previously determined
proper motions in Table~\ref{tab:vla_previous} with differences
between the previous and our proper motion components expressed in
units of the combined-in-quadrature uncertainties.

Because of their high precision, our results are valuable for testing
the reliability of the published error bars of previous
interferometric measurements.  We have thirteen sources that have also
been observed by other studies: five in a one-baseline experiment at
Jodrell Bank \citep{las82}, four in a two-baseline experiment at
Jodrell Bank \citep{hla93}, and four in a previous program at the VLA
\citep{fgml97}.  Except for two results that deviated by more than
$3\sigma$, the proper motions in right ascension agree quite well,
whereas only five of thirteen pulsars agree to better than $2 \sigma$
in declination, suggesting an underestimate of the declination proper
motion uncertainties in previous measurements.  It is possible that
the ionosphere and unaccounted structures of sources near the pulsars
in the Jodrell Bank proper motions contributed to additional
uncertainty.  The ionosphere will tend to increase uncertainties in
the declination more than the right ascension since most observations
are scheduled near transit, the time when the gradient in the
ionosphere's strength is largely in declination.  Our use of 1452~MHz
reduced the ionosphere's effect by a factor of 12.7 over the
previous 408~MHz observations at Jodrell Bank.

\section{Distance uncertainties \label{sec:dist}} 

If a pulsar's distance is known, the observed proper motion can be
reduced to a physical transverse velocity; hence distance measurements
are very important for kinematic studies of pulsars.

A pulsar's distance can be estimated by combining the line-of-sight
electron column density, or dispersion measure (DM), with a model of
the Galactic electron distribution.  Normally, uncertainty in the DM
can be ignored, so the precision of this method depends only on the
Galactic electron model.  For the last decade, the standard model has
been that of Taylor and Cordes (1993; hereafter T\&C)\nocite{tc93}.
\footnote{Although new models (G\'{o}mez et al. 2001 and 
Cordes \& Lazio 2002) have
recently been introduced, we continue to use the T\&C model for
consistency with earlier results.} Except for a few pulsars associated
with objects at known distance, such as globular clusters, annual
parallax is the only model-independent method to obtain the distance to
a pulsar.  Because most parallax measurements were made after the
development of the T\&C model, and were not used in the development of
the model, they provide a uniquely powerful test of the model. A
comparison was made by \cite{bbgt02}, who found that the model could
predict the DM to a given distance to about 40\% accuracy.  Although
this fractional precision was determined from pulsars within about a
kiloparsec of the Sun, we have assumed the same precision at greater
distances.  

\begin{figure}[tp]
\begin{center}
\resizebox{5in}{!}{\includegraphics{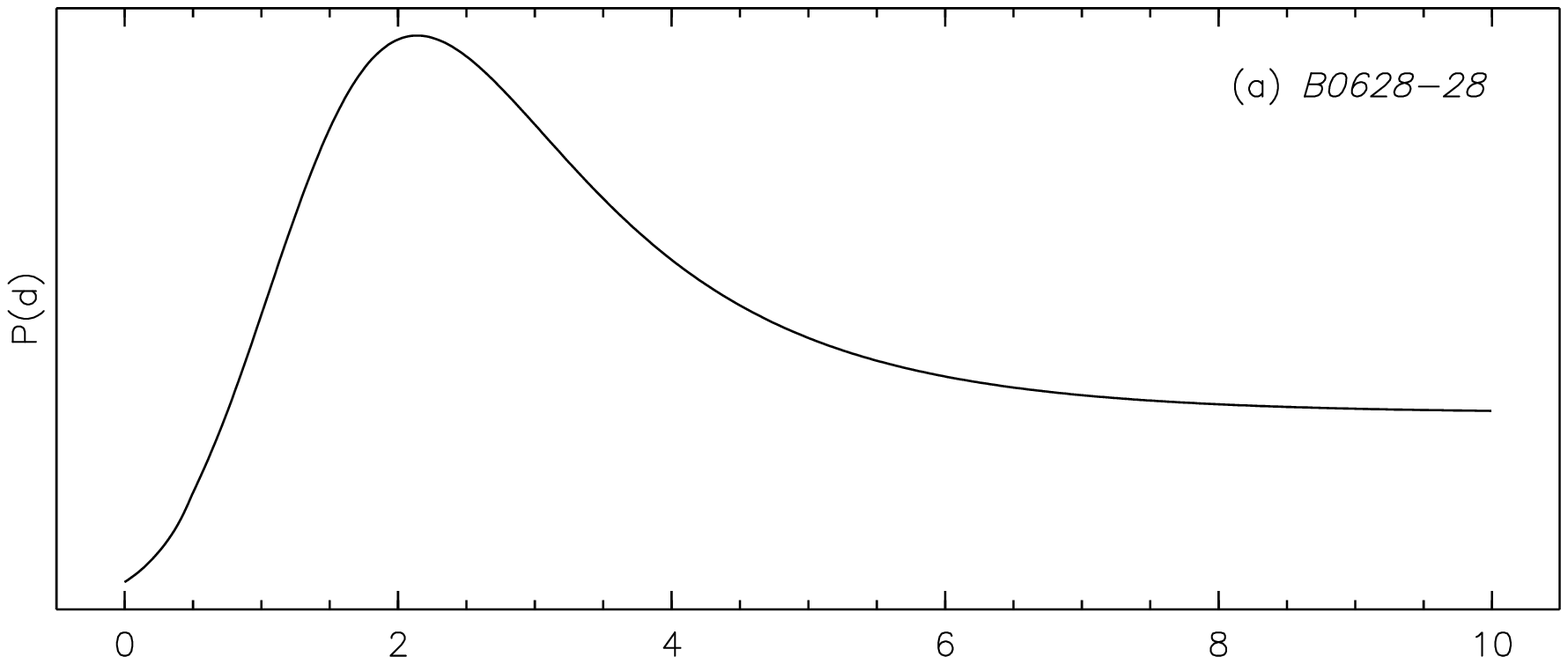}} \\
\resizebox{5in}{!}{\includegraphics{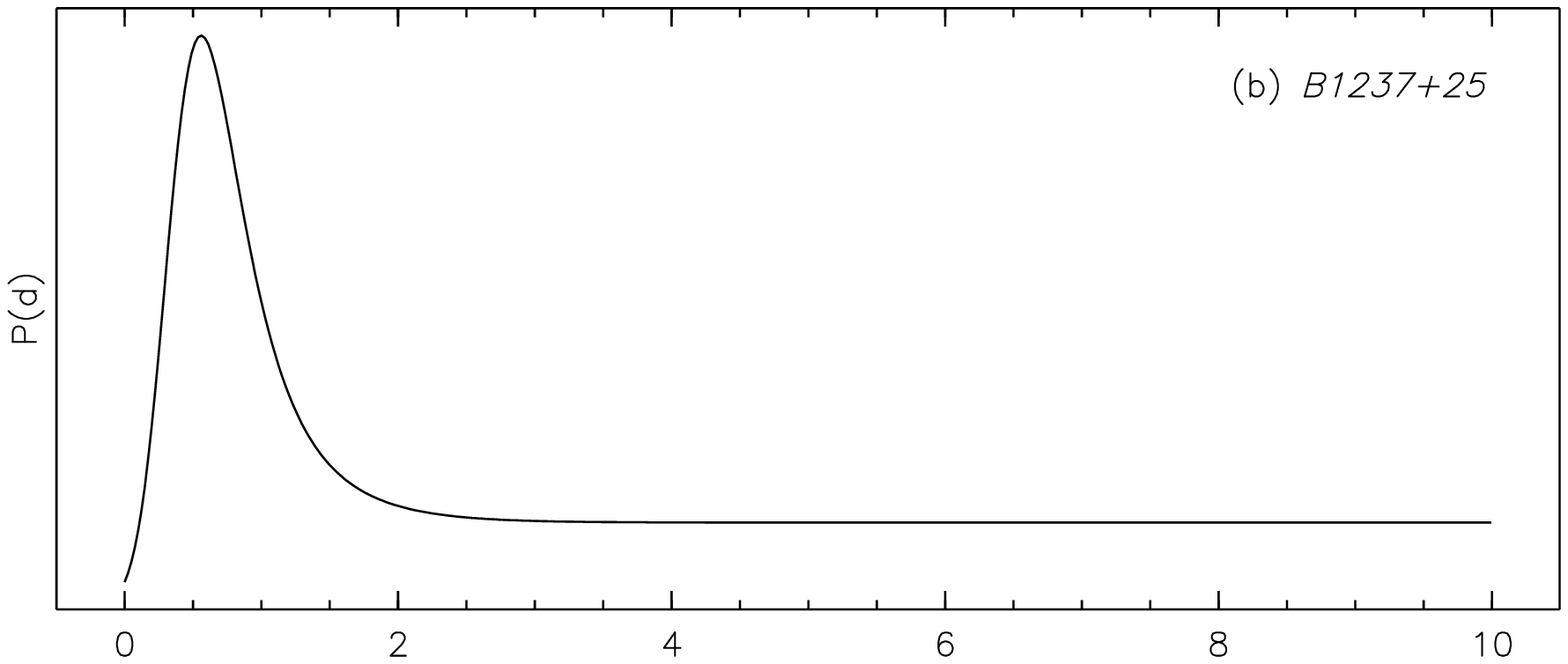}} \\
\resizebox{5in}{!}{\includegraphics{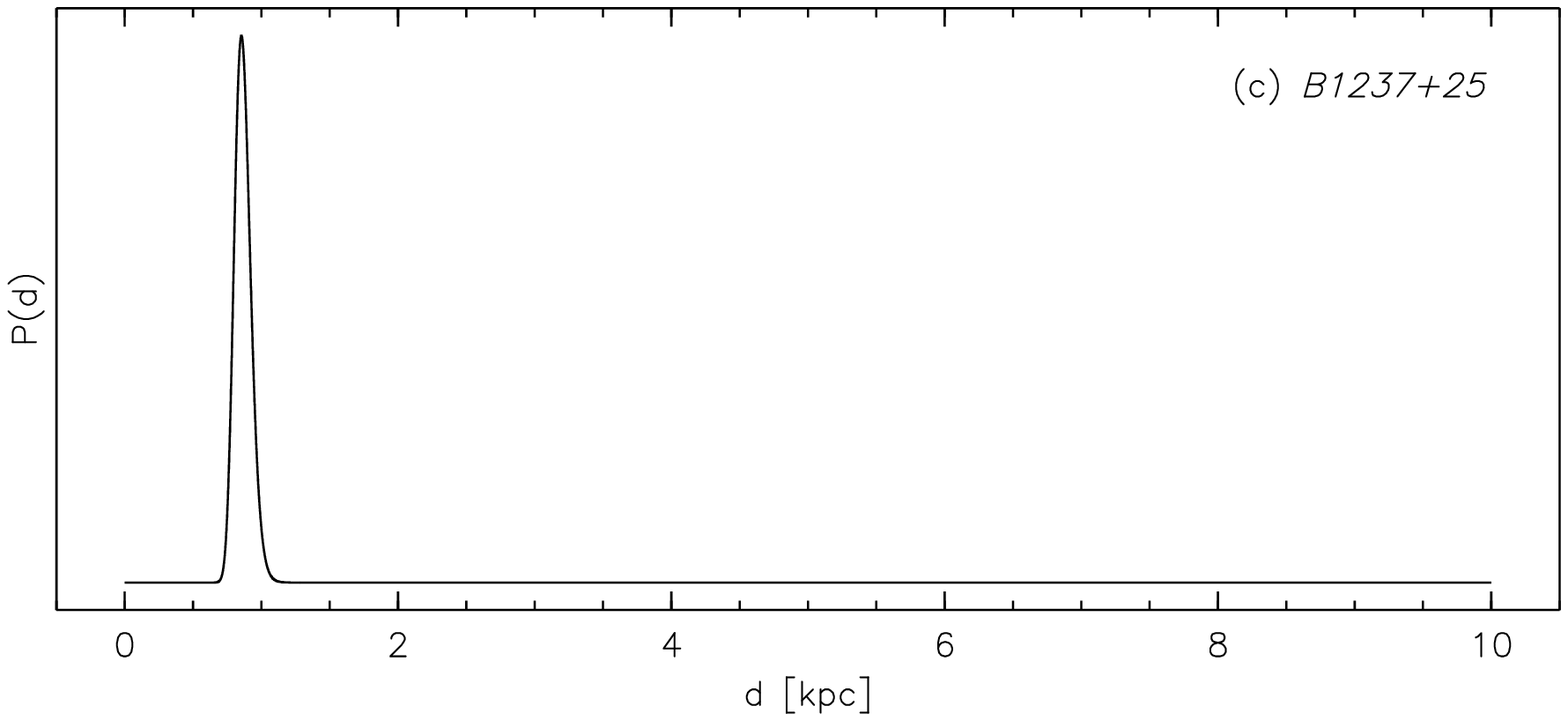}} 
\end{center}
\caption{
\label{fig:distpdf}
Examples of different distance PDFs, demonstrating their non-Gaussianity.
The DM-derived distance PDFs for B0628-28 and B1237+25 are shown in
panels (a) and (b) respectively.  Note the finite probability density even
at very large distances.  The much more constraining PDF derived for 
B1237+25 from parallax is shown in panel (c).  These PDFs are 
normalized so that their integral is 1.
}
\end{figure}

Understanding the statistics of distance measurements is important
when modeling populations, since incorrect uncertainty estimates can
bias results.  For example, a sample with excessively large estimates
of uncertainty will be biased towards greater distance. Further, in
almost all cases, the non-uniform model distribution of electrons
produces significantly non-Gaussian distance uncertainties.  At the
extreme are pulsars with DMs comparable to the
model's total electron column density in a particular distance.  In
these cases the pulsars' distances are unbounded by the DM
measurement.  Examples of distance uncertainties from both dispersion
and parallax measurements are shown in Fig.~\ref{fig:distpdf}.

\section{Discussion \label{sec:disc}}

Given a sample of pulsars with well-measured proper motions, we can
address a number of questions about pulsar kinematics.  However, even
with 28 pulsars our sample is small.  We considered adding earlier
proper motion measurements, most with poorer precision, but as
described above we have found that the error bars assigned to these
measurements are not well understood.  We were, however, able to
supplement our sample with six additional sources younger than 20~Myr, 
whose proper
motions were reliably measured by Brisken et al.\ (2002), using VLBI.
Table~\ref{tab:B2002} contains a list of these pulsars and some of
their properties.

\subsection{The birth velocity distribution \label{sec:vel}}

Extraction of the pulsar birth velocity distribution from a selected
sample of proper motions of middle-aged pulsars with poorly known
distances is a task riddled with bias, uncertainty, and confusion.
Nevertheless, the result is crucial to such questions as the fraction
of pulsars that escape the Solar neighborhood (and in turn the local
birthrate, the supernova rate, the rate of enrichment of the
interstellar medium, and the minimum mass for a star to undergo a supernova
explosion),
the fraction of newborn neutron stars retained in globular clusters,
and the dynamics of supernova collapse.  Hence we follow many other
authors in tackling the question, while cautioning that much work
remains and our results should not be overinterpreted.  In particular,
although our sample was chosen specifically to reduce sample bias, we
believe the sample size is still too small to effectively measure and
remove the residual bias that undoubtedly remains.

Sample bias entered the problem initially in the pulsar discovery
surveys.  Searches are obviously dependent on the pulsar flux density
and hence on luminosity and distance, but they are more subtly biased
even for pulsars with the same luminosity and distance.  For example,
the sensitivity depends (in a generally complex way) on DM, galactic
coordinates, declination, pulse period, and pulse shape.  All of these
quantities may be correlated with velocity.  For example, if proper
motion and spin axis are correlated, then pulsars that have moved far
from the galactic plane will be detected with less efficiency, or if
luminosity is related to velocity then the resultant velocity
distribution will be biased.  Age is almost certainly correlated with
distance from the galactic plane, meaning searches targeting the
Galactic plane will inherently bias the sample of known pulsars.
Pulsars moving quickly away from the Galactic plane spend less time
within their detection radius.  As pulsars get farther away their
motion tends toward purely radial.  Additionally, the Galactic
gravitational potential significantly accelerates a pulsar after a
period of about $10^7$~years.

Bias gets compounded when proper motion samples are chosen.  It is
most straightforward to measure proper motions of brighter objects,
which tend to be even closer than the typical known pulsars. Since the
highest speed pulsars leave the local sample volume more quickly than
the slower pulsars the resultant mean velocity is lowered.

Ideally, proper motion studies would sample a very large volume and be
limited to pulsars that are young enough to minimize loss from the
sample or deceleration in the Galactic potential.  Discovery and
measurement limitations and the small number of very young pulsars
make such an ideal study impossible, leading many authors
\citep{hp97,acc02,lbh97} to attempt to measure or otherwise estimate
the biases and correct the sample.  We have taken a different
approach, trying to reduce the size of the biases with more accurate
measurements that allowed a larger sample volume.  However, the
tradeoff is that, in our judgment, the resulting sample size is too
small to allow sophisticated analysis of the remaining bias.

We have addressed the problem of Galactic acceleration, and further
reduced the problem of sample incompleteness, by limiting our analysis
to the single velocity component perpendicular to the line of sight and
parallel to the Galactic plane: $v_l$ is the product of distance, $D$,
and the corresponding component of proper motion, $\mu_l$. In our
analysis, we first tabulate for each pulsar $P(v_l)$, the PDF
for $v_l$.  This calculation is complicated by the
distance dependence on the Galactic rotation correction,
$\Delta_{v_l}(D)$, and assumes the form 
\begin{equation} P(v_l) =
\int_0^\infty d D \int_{-\infty}^\infty d \mu_l \,
\delta \left( v_l - \left(
D\,\mu_l - \Delta_{v_l}(D)\, \right) \right) P(D) \, P(\mu_l).
\end{equation} 
In practice, this is integrated numerically from
tabulated $P(D)$ and $P(\mu_l)$ PDFs.

\begin{figure}[tp]
\begin{center}
\resizebox{5.4in}{!}{\includegraphics{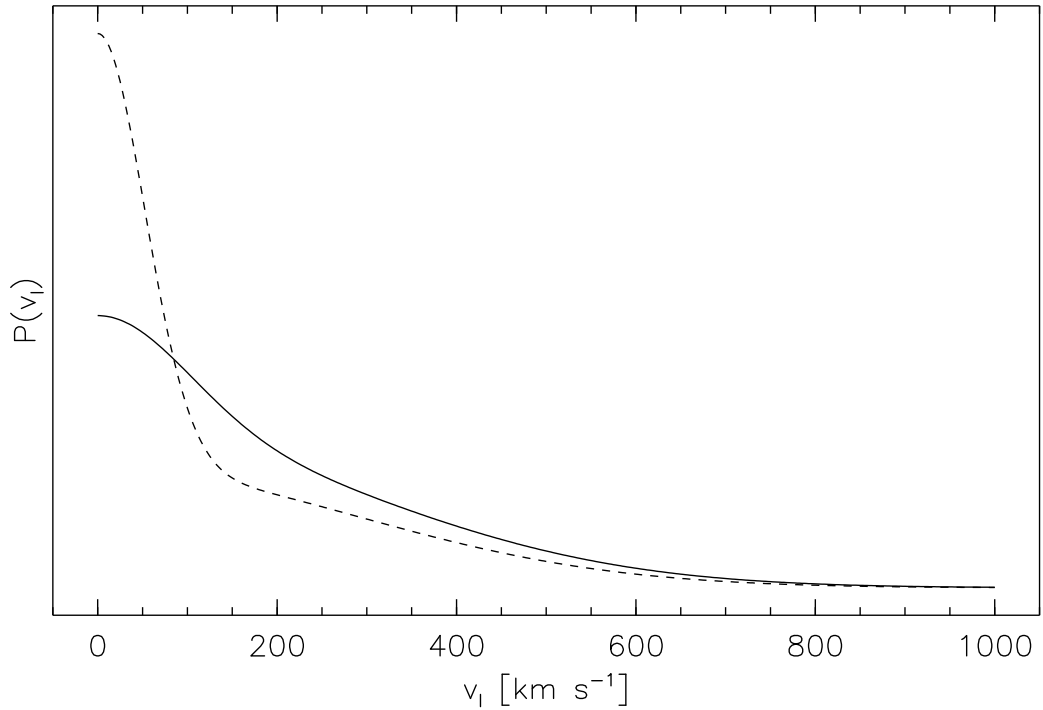}}
\end{center}
\caption{
\label{fig:vel}
1-dimensional best fit velocity PDFs from our proper motion study (solid) and
from ACC (dashed).
}
\end{figure}

With this set of PDFs different velocity models can be tested.  The
likelihood, $L$, that a model $P(v_l; \vec{m})$ with model parameters
$\vec{m}$, explains the measurements, $P_i(v_l)$ is: 
\begin{equation} \label{eqn:like} 
L = \prod_i \int dv_l\,  P_i(v_l) \, P(v_l;\vec{m}). 
\end{equation} 
Comparing different model forms, especially
those with different numbers of parameters, is difficult to do in an
unbiased manner, we will use the same model form that was used by
\citet{acc02} (hereafter ACC). This model contained three parameters
describing the widths of two zero-centered Gaussian distributions, and
the fraction of pulsars in the first:
\begin{equation} \label{eqn:model} P(v_l; \sigma_1, \sigma_2, w_1) =
\frac{w_1}{\sqrt{2 \pi \sigma_1^2}} \exp(-v_l^2 / 2\sigma_1^2) +
\frac{1-w_1}{\sqrt{2 \pi \sigma_2^2}} \exp(-v_l^2 / 2\sigma_2^2).
\end{equation} 
This is the analog of Eqn.~1 in ACC for use with a
single component of the velocity, hence the velocity dispersions,
$\sigma_{v_i}$ in ACC is a factor of $\sqrt{3}$ larger than the
corresponding value, $\sigma_i$, in Eqn.~\ref{eqn:model}. Maximizing
the likelihood (Eqn.~\ref{eqn:like}) results in a best fit with
$\sigma_1 = 99$~km~s$^{-1}$, $\sigma_2 = 294$~km~s$^{-1}$, and $w_1 =
0.20$.  These model parameters are preferred to those of ACC ($\sigma_1
= 52$~km~s$^{-1}$, $\sigma_2 = 289$~km~s$^{-1}$, $w_1 = 0.4$) by
an odds ratio of 7.7 (see ACC).  
A plot of the resultant 1-dimensional velocity PDFs for
both our fit and that of ACC is shown in Fig.~\ref{fig:vel}.

\subsection{Pulsar birth locations\label{sec:birth}}

Based on the low scale height of massive O and B stars,
it is conventionally assumed that most or all neutron stars recently
formed in the Milky Way were born very near the Galactic plane.  However,
Harrison et al. (1993) found that 17\% of the young pulsars in their sample
were falling towards the plane, perhaps indicating a second group of
pulsar progenitors with large scale height.  As pulsars born out of
the plane only have a 50\% chance of being born with a velocity pointing
towards the plane, and as many of these will rapidly pass through the plane
(a pulsar born at 150~pc above the plane moving towards the plane with
150~km~s$^{-1}$ only takes one million years to reach the plane) and then appear
to be moving away from the plane,  the work of Harrison et al., if confirmed,
would imply a very large high birth-height population.
One original motivation
for our work was, therefore, to determine better the size of
the population of infalling pulsars.

Old pulsars may be moving towards the plane not because they
were born with velocity vectors pointing towards the plane,
but rather because the Galactic potential has over course
time caused their infall.  Pulsars older than about 20~Myr
will have undergone a significant fraction of an oscillation
about the Galactic plane, and therefore could potentially
contaminate any study of the birth heights of pulsars.

A geometric method
to estimate a pulsar's age is to divide the angular
distance traversed during the pulsar's lifetime by its proper motion.
Again, for this age estimate to be reliable, its true age 
must be less than about 10~Myr.  This kinetic age,
$\tau_{\mathrm{kin}}$, is given as
\begin{equation}
\tau_{\mathrm{kin}} = \frac{D \sin{b} - z_0}{\mu_b D \cos{b} + v_r \sin{b}},
\end{equation}
where $D$ is the pulsar's distance, $z_0$ is its birth height, and
$v_r$ is the pulsar's radial velocity.  The birth height and radial
velocity cannot be measured so an initial distribution must be
assumed.  A birth height distribution with scale height $\sigma_{z_0} =
150$~pc is assumed.  The pulsar birth height distribution is not well known.  
The distribution chosen here is chosen as a compromise between a modeled
pulsar birth height of 
175~pc (Stollman, 1987), and the supernova scale height, $\sim 100$~pc
(based on Green, 2001).  Note that both of these estimates are made 
without pulsar kinematics statistics.  It should be noted that this
scale height is consistent with that derived by the simulations of 
ACC ($z_0=160\pm40$~pc).
The radial velocity distribution is a 1-D
component of the full velocity, analogous to what was measured in
\S~\ref{sec:vel}.  The radial velocity requires a distance dependent
correction, $\Delta v_r(D)$, due to differential Galactic rotation and
solar motion.
Although we are aware that various projection effects will cause the
radial velocity distribution to be different than the velocity
distribution measured above, we will use that distribution.  The PDF
for kinetic age can then be computed:
\begin{eqnarray} \label{eqn:taukin}
P(\tau_\mathrm{kin}) & = & 
	\int \! d z_0 \int \! d D \int \! d v_r \int \! d \mu_b \,
\delta \left(
        \tau_\mathrm{kin} - \frac{D \sin(b) - z_0}{\mu_b D \cos(b) + 
		\left(v_r + \Delta v_r(D)\right)\sin(b)}
       \right) \times \nonumber \\
              &   & P_{z_0}(z_0|\sigma_{z_0}) P_D(D|DM) P_{v_r}(v_r)
P_{\mu_b}(\mu_b|\mu_{b, \mathrm{meas}}, \sigma_{\mu_b}).
\end{eqnarray}

A negative kinetic age would imply that the pulsar is falling toward the
plane. 
The probability that a pulsar is falling towards the plane,
$P_{\mathrm{fall}}$ can be computed:
\begin{equation}
P_{\mathrm{fall}} = \int_{-\infty}^{\,0} \! d \tau_{\mathrm{kin}} 
\ P_{\tau_{\mathrm{kin}}}(\tau_\mathrm{kin}).
\end{equation}

From a sample of pulsars with calculated $P_{\mathrm{fall}}$ we wish to 
calculate the probability $P(f)$ that a fraction
$f$ of young pulsars are falling based on the set of $P_\mathrm{fall}$ values
for pulsars with ages less than 20~Myr.  Although significant acceleration
could occur within 20~Myr, less than one quarter of an orbit would have
completed by this time, which maintains the sign of $\tau_{\mathrm{kin}}$
\begin{eqnarray}
P(f) & = & P(f|\{P_\mathrm{fall,i}\}) \nonumber\\
     & \propto & P(\{P_\mathrm{fall,i}\}|f) P_{\mathrm{prior}}(f),
\end{eqnarray}
where Bayes theorem was again used.  A flat prior PDF, $P_{\mathrm{prior}}(f)$,
is assumed.  Since all of the data are independent this can be
expanded as a product of single pulsar probabilities:
\begin{equation}
P(f) \propto \prod_i P(P_\mathrm{fall,i}|f),
\end{equation}
where the conditional PDF has the form
\begin{equation}
P(P_\mathrm{fall}|f) = 2P_\mathrm{fall}f + 2(1-P_\mathrm{fall})(1-f).
\end{equation}
Note that integrating this over $0 \le P_\mathrm{fall} \le 1$ yields
1, regardless of the value of $f$.

\begin{figure}[tp]
\begin{center}
\resizebox{5.4in}{!}{\includegraphics{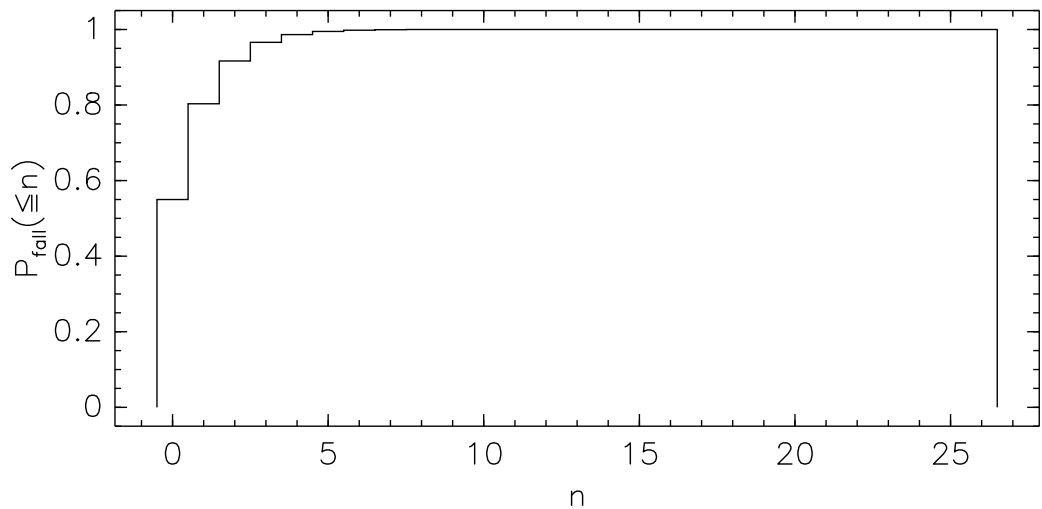}}
\end{center}
\caption[Pulsars falling toward the plane] {
\label{fig:vel_falling}
Cumulative probability distribution for the number of young
($\tau_{\mathrm{char}} < 20$ Myr)
pulsars out of the 26 pulsar sample falling toward the plane.  }
\end{figure}

We used a sample of 26 pulsars (20 from this study, and 6 from Brisken et~al. (2002))
with characteristic age less than 20~Myr.
$P(f)$ was computed on a grid of values between 0 and 1, and is
plotted in Fig.~\ref{fig:vel_falling}.  We conclude that, with 68\%
confidence, fewer than 6\% of young pulsars are falling towards the
plane (fewer than 11\% at 95\% confidence).  Contrary to earlier
studies, we find that a population of pulsars with large birth height
is not required.  

\subsection{Timing ages and braking indexes \label{sec:brake}}

Timing properties can be used to estimate the age of a pulsar.  The basic 
assumption that a pulsar's spin frequency, $\nu = 1/P$, evolves via 
power-law decay, $\dot{\nu} \propto -\nu^n$, leads to an age,
\begin{equation}\label{eqn:age}
\tau = \frac{P}{(n-1)\dot{P}} \left[ 1 - \left(\frac{P_0}{P}\right)^{n-1}\right],
\end{equation}
where $P$ and $\dot{P}$ are the pulsar's spin period and spin period 
time derivative, respectively.  The pulsar's initial spin period, $P_0$,
is generally assumed to be much smaller than its current period.  
However, the 39~ms pulsar B1951+32, associated with supernova remnant CTB80, 
has $P_0 = 27 \pm 6$~ms (Migliazzo et al. 2002).  Likewise,
the 143~ms pulsar J0538+2817 must have a $P_0 \gtrsim 130$~ms based on the 
likely association with S147 (Romani \& Ng 2003).  These examples
suggest that
$P_0 \ll P$ is not strictly true.
The braking index $n$ depends on the dominant
torque acting on the pulsar.  It is believed that electromagnetic dipole
radiation, with $n=3$, dominates for most pulsars.  The characteristic age,
$\tau_{\mathrm{char}}$, is thus defined by Eqn.~\ref{eqn:age} with
this value of $n$ and with the assumption that $P_0 \ll P$:
\begin{equation} \label{eqn:tchar}
\tau_{\mathrm{char}} = \frac{P}{2\dot{P}}.
\end{equation}
This value is generally used to estimate pulsar ages.

\begin{figure}[tp]
\begin{center}
\resizebox{5.4in}{!}{\includegraphics{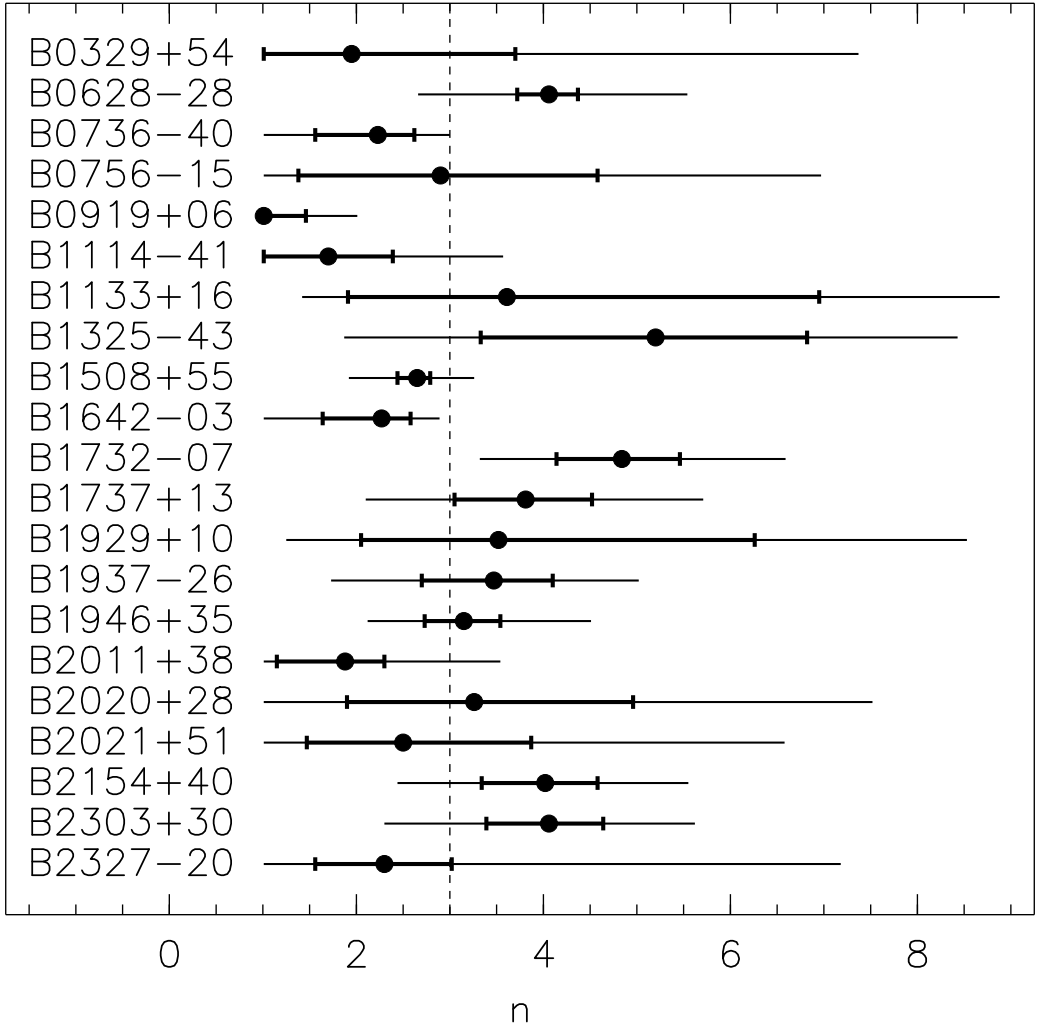}}
\end{center}
\caption[n PDF] {
\label{fig:nvalues}
The braking indexes, $n$, for the 21 pulsars younger than 10~Myr.  
The most likely value of
$n$ for each pulsar is shown by a dot.  The most compact 68\% confidence
intervals are denoted by the thick horizontal lines with vertical bars.  The
most compact 95\% confidence intervals are denoted by the thin horizontal 
lines.  The vertical dashed line indicates the nominal value of 3.
}
\end{figure}

A pulsar's kinetic age PDF can be used to
determine its braking index PDF.  Since the initial periods of
none of the pulsars in the sample are known, a scale invarient PDF for the
initial period was used with a lower cutoff of 5~ms.  The smaller of
the pulsar's period or 200~ms was used as the upper cutoff.  The results are
not very sensitive to these cutoffs.  It is implicitly assumed that the
braking index of a pulsar remains constant throughout its life.
The desired PDF can be expressed as
\begin{equation}
P_n(n) = \int \! d P_0 \int \! d \tau \, \delta(n - 
	n(P, \dot{P}, P_0, \tau)) \, P_{\tau}(\tau) \, P_{P_0}(P_0),
\end{equation}
where $n(P, \dot{P}, P_0, \tau)$ is Eqn.~\ref{eqn:age} reexpressed for 
the value of $n$, and $\tau$ is the true age of the pulsar.  Here we use the
kinetic age for the true 
age and its PDF, $P_{\tau}(\tau)$ is derived in Eqn.~\ref{eqn:taukin}.
The braking index PDFs for 21 pulsars (16 from this paper and 5 from Brisken et~al., 2002) with $\tau_{\mathrm{char}} < 10$~Myr 
and $|b| > 1^\circ$ are shown schematically in Fig.~\ref{fig:nvalues}.

The distribution of pulsar braking indexes can be derived from a set of
braking index PDFs.  In this analysis we assume the form of the braking
index distribution is Gaussian with two model parameters, the mean, $n_0$,
and the width, $\sigma_n$.  The distributions
are truncated below $n = 1$.  Flat priors are used for
$n_0$ and $\sigma_n$.  This
analysis proceeded analogously to Eqn.~\ref{eqn:like}.
The model parameters yielding the greatest likelihood are $n_0 = 2.98$ and
$\sigma_n = 0.82$.  This width is prefered over the most likely zero-width 
distribution, that with $n_0 = 2.64$, by
an odds ratio of 403.  This strongly suggests 
that there is an intrinsic width to the
distribution of braking indexes, as expected given the handful of direct
measurements of the braking indexes of young pulsars
(see Table~\ref{tab:brake}).  It is not certain that these values are 
representative of all pulsars.  The magnitude of the width suggests that
the characteristic age, $\tau_{\mathrm{char}}$,
can only be trusted to about 40\%.

\acknowledgments SET is supported by NSF grant AST-0098343.  The graduate work
of WFB was partially funded by an NSF graduate student fellowship.  Special
thanks go to Mike Nolta for the creation of the Biggles plotting package,
which was used exclusively in this publication.  We thank the anonymous
referee who provided useful feedback.

\bibliographystyle{apj}

\clearpage

\begin{deluxetable}{lllllr}
\tablecaption{Pulsars observed\label{tab:obs}}
\tablehead{
\colhead{Pulsar\tablenotemark{1}} & \colhead{Observed} & 
\colhead{Flux} & \colhead{Period} &
\colhead{$\tau_{\mathrm{char}}$} & \colhead{Number}\\
&\colhead{Epochs\tablenotemark{2}} & Density\tablenotemark{3}&&&\colhead{of Ref.}\\
&&\colhead{(mJy)}&
\colhead{(s)}&\colhead{(Myr)}&\colhead{Sources}}
\startdata
B0011+47  & 94g 98bg 99g      & 5   & 1.241 & 34.9 &  21  \\ 
B0149--16 & 86 92g 95g 99g    & 1.6 & 0.833 & 10.2 &  17  \\ 
B0628--28 & 86 92 95 99       & 2   & 1.244 & 2.77 &  17  \\ 
B0736--40 & 86 95 99          & 50  & 0.375 & 3.68 &   7  \\ 
B0756--15 & 94g 98g 99g       & 0.8 & 0.682 & 6.69 &  10  \\ 
B0835--41 & 94g 98 99         & 20  & 0.752 & 3.36 &  12  \\ 
B0919+06  & 86 92g 95 98 99g  & 6   & 0.431 & 0.50 &  11  \\ 
B1039--19 & 94g 98g 99g       & 1.4 & 1.386 & 23.3 &   5  \\ 
B1114--41 & 92 95 99          & 3   & 0.943 & 1.88 &   4  \\ 
B1237+25  & 92 95 98 99g      & 15  & 1.382 & 22.8 &   9  \\ 
B1325--43 & 94g 98bg          & 2   & 0.533 & 2.80 &   9  \\ 
B1508+55  & 92g 95 99g        & 8   & 0.740 & 2.34 &  14  \\ 
B1534+12  & 94g 98bg 99       & 2   & 0.038 & 247  &  16  \\ 
B1540--06 & 84 92g 94g 98bg 99& 2   & 0.709 & 12.7 &  10  \\ 
B1541+09  & 92 95 99          & 5.9 & 0.748 & 27.4 &  25  \\ 
B1552--31 & 94g 99g           & 1.4 & 0.518 & 132  &  12  \\ 
B1642--03 & 92 95 99          & 21  & 0.388 & 3.45 &   7  \\ 
B1718--02 & 94g 99g           & 1   & 0.478 & 87.0 &   6  \\ 
B1732--07 & 92g 99g           & 1.7 & 0.419 & 5.47 &   8  \\ 
B1737+13  & 92g 99g           & 3.9 & 0.803 & 8.75 &  11  \\ 
B1937--26 & 94 98 99g         & 3   & 0.403 & 6.68 &   7  \\ 
B1943--29 & 92 99g            & 0.8 & 0.959 & 10.2 &  12  \\ 
B1946+35  & 84 92 95 99       & 8.3 & 0.717 & 1.61 &  21  \\ 
B2011+38  & 94 98 99          & 6.4 & 0.230 & 0.41 &  12  \\ 
B2106+44  & 92g 99g 01        & 5.4 & 0.415 & 76.3 &   9  \\ 
B2154+40  & 92g 95g 99        & 17  & 1.525 & 7.05 &  22  \\ 
B2303+30  & 92g 95g 99g       & 2.2 & 1.576 & 8.62 &  11  \\ 
B2327--20 & 94g 98bg 99       & 3   & 1.644 & 5.63 &  13  \\ 
\enddata
\tablenotetext{1}{Period, period derivative, and flux density are from the catalog
of \citet{tmlc95}.}
\tablenotetext{2}{last two digits of years of observation. Suffix `b' indicates 
that some data from the A- to B-array reconfiguration was used.
Suffix `g' indicates pulsar gating.}
\tablenotetext{3}{at 1400~MHz}
\end{deluxetable}

\begin{deluxetable}{lrrrrrrrrrr}
\rotate
\tabletypesize{\scriptsize}
\tablecaption{Measured parameters\tablenotemark{a}\label{tab:res}}
\tablehead{
\colhead{Pulsar} & \colhead{Right Asc.} & \colhead{Dec.} & \colhead{l} &
\colhead{b} & \colhead{$\mu_\alpha\,\cos\delta$} & \colhead{$\mu_\delta$} & \colhead{Cov.} &
\colhead{$\mu_l\,\cos b$} & \colhead{$\mu_b$} & \colhead{Cov.}}
\startdata
B0011+47  & 00:14:17.757  &   47:46:33.29  & 116.49 & --14.63 & 19.3 $\pm$ 1.8\phn    & --19.7 $\pm$ 1.5\phn  & 0.54   &   16.1  $\pm$  1.9\phn  & --22.3 $\pm$  1.4\phn  &   0.48 \\
B0149--16 & 01:52:10.858  & --16:37:53.22  & 179.30 & --72.45 & 3.1 $\pm$ 1.2\phn     & --27.2 $\pm$ 2.0\phn  & --0.12 &   23.0  $\pm$  1.9\phn  &  --14.8 $\pm$  1.6\phn  & --0.47 \\
B0628--28 & 06:30:49.433  & --28:34:42.91  & 236.95 & --16.75 & --44.6 $\pm$ 0.9\phn  & 19.5 $\pm$ 2.2\phn    & 0.05   &  --34.9  $\pm$  2.1\phn  & --33.9 $\pm$  1.2\phn  & --0.57 \\
B0736--40 & 07:38:32.329  & --40:42:40.94  & 254.19 &  --9.19 & --14.0 $\pm$ 1.2\phn  & 12.8 $\pm$ 2.1\phn    & --0.54 &  --17.9  $\pm$  2.2\phn  &  --6.3 $\pm$  1.0\phn  & --0.25 \\
B0756--15 & 07:58:29.109  & --15:28:09.78  & 234.46 &    7.22 & 1.2 $\pm$ 4.0\phn     & 3.6 $\pm$ 5.6\phn     & --0.31 &   --2.4  $\pm$  5.7\phn  &    2.9 $\pm$  3.7\phn  & --0.15 \\
B0835--41 & 08:37:21.266  & --41:35:15.05  & 260.90 &  --0.33 & --2.3 $\pm$ 1.8\phn   & --17.8 $\pm$ 3.3\phn  & --0.59 &   12.8  $\pm$  3.4\phn  & --12.6 $\pm$  1.7\phn  & --0.48 \\
B0919+06  & 09:22:14.008  &   06:38:22.70  & 225.42 &   36.39 & 18.8 $\pm$ 0.9\phn    & 86.4 $\pm$ 0.7\phn    & 0.05   &  --66.5  $\pm$  0.8\phn  &   58.4 $\pm$  0.8\phn  &   0.09 \\
B1039--19 & 10:41:36.201  & --19:42:13.44  & 265.59 &   33.59 & --1.1 $\pm$ 2.6\phn   & 14.4 $\pm$ 4.5\phn    & 0.15   &   --9.2  $\pm$  3.1\phn  &   11.2 $\pm$  4.2\phn  & --0.45 \\
B1114--41 & 11:16:43.086  & --41:22:43.96  & 284.45 &   18.06 & --1.4 $\pm$ 4.8\phn   & 7.1 $\pm$ 20.3    & 0.05   &   --3.9  $\pm$  8.6\phn  &    6.1 $\pm$ 19.0  & --0.80 \\
B1237+25  & 12:39:40.361  &   24:53:49.96  & 252.44 &   86.54 & --104.5 $\pm$ 1.1\phn & 49.4 $\pm$ 1.4\phn    & 0.05   & --105.6  $\pm$  1.2\phn  & --46.8 $\pm$  1.3\phn  & --0.20 \\
B1325--43 & 13:28:06.432  & --43:57:44.12  & 309.87 &   18.41 & 3.3 $\pm$ 6.6\phn     & 53.6 $\pm$ 22.6   & 0.18   &   11.3  $\pm$  7.9\phn  &   52.5 $\pm$ 22.2  &   0.54 \\
B1508+55  & 15:09:25.646  &   55:31:32.53  &  91.32 &   52.28 & --70.6 $\pm$ 1.6\phn  & --68.8 $\pm$ 1.2\phn  & 0.22   &  --16.7  $\pm$  1.2\phn  &   97.2 $\pm$  1.6\phn  &   0.17 \\
B1534+12  & 15:37:09.960  &   11:55:55.57  &  19.84 &   48.34 & --7.6 $\pm$ 9.3\phn   & --31.6 $\pm$ 10.1 & 0.03   &  --31.5  $\pm$ 10.0  &  --7.9 $\pm$  9.4\phn  &   0.04 \\
B1540--06 & 15:43:30.167  & --06:20:45.42  &   0.56 &   36.60 & --17.4 $\pm$ 2.4\phn  & --3.6 $\pm$ 2.7\phn   & 0.42   &  --14.1  $\pm$  3.0\phn  &   10.8 $\pm$  1.9\phn  &   0.02 \\
B1541+09  & 15:43:38.836  &   09:29:16.41  &  17.81 &   45.77 & --7.3 $\pm$ 1.0\phn   & --4.0 $\pm$ 1.0\phn   & --0.12 &   --7.0  $\pm$  1.0\phn  &    4.4 $\pm$  1.0\phn  &   0.11 \\
B1552--31 & 15:55:17.946  & --31:34:20.16  & 342.69 &   16.75 & 60.6 $\pm$ 19.3   & --77.3 $\pm$ 50.6 & --0.37 &   --6.6  $\pm$ 31.5  & --98.0 $\pm$ 44.0  &   0.75 \\
B1642--03 & 16:45:02.051  & --03:17:57.93  &  14.11 &   26.06 & --3.7 $\pm$ 1.5\phn   & 30.0 $\pm$ 1.6\phn    & 0.53   &   23.3  $\pm$  1.9\phn  &   19.2 $\pm$  1.1\phn  & --0.22 \\
B1718--02 & 17:20:57.257  & --02:12:24.15  &  20.13 &   18.93 & --0.8 $\pm$ 3.7\phn   & --25.7 $\pm$ 4.5\phn  & 0.18   &  --22.7  $\pm$  4.7\phn  & --12.0 $\pm$  3.6\phn  &   0.08 \\
B1732--07 & 17:35:04.965  & --07:24:52.29  &  17.27 &   13.28 & --2.4 $\pm$ 1.7\phn   & 28.4 $\pm$ 2.5\phn    & 0.01   &   23.3  $\pm$  2.4\phn  &   16.4 $\pm$  2.0\phn  &   0.31 \\
B1737+13  & 17:40:07.332  &   13:11:56.64  &  37.08 &   21.67 & --21.5 $\pm$ 2.2\phn  & --19.7 $\pm$ 2.2\phn  & --0.09 &  --26.7  $\pm$  2.1\phn  &   11.5 $\pm$  2.3\phn  &   0.06 \\
B1937--26 & 19:41:00.405  & --26:02:05.98  &  13.90 & --21.82 & 12.1 $\pm$ 2.4\phn    & --9.9 $\pm$ 3.8\phn   & 0.01   &   --5.0  $\pm$  3.6\phn  & --14.8 $\pm$  2.6\phn  &   0.27 \\
B1943--29 & 19:46:51.732  & --29:13:47.03  &  11.10 & --24.12 & 18.6 $\pm$ 8.9\phn    & --32.8 $\pm$ 20.3 & --0.46 &  -25.1  $\pm$ 18.1  & --28.2 $\pm$ 12.8  &   0.72 \\
B1946+35  & 19:48:25.000  &   35:40:11.04  &  70.70 &    5.04 & --12.6 $\pm$ 0.6\phn  & 0.7 $\pm$ 0.6\phn     & --0.11 &   --6.4  $\pm$  0.8\phn  &   12.4 $\pm$  0.9\phn  &   0.12 \\
B2011+38  & 20:13:10.355  &   38:45:43.17  &  75.93 &    2.47 & --32.1 $\pm$ 1.7\phn  & --24.9 $\pm$ 2.3\phn  & --0.49 &  --38.5  $\pm$  1.7\phn  &   13.1 $\pm$  2.3\phn  &   0.47 \\
B2106+44  & 21:08:20.481  &   44:41:48.81  &  86.90 &  --2.01 & 3.5 $\pm$ 1.3\phn     & 1.4 $\pm$ 1.4\phn     & 0.05   &    3.4  $\pm$  1.4\phn  &  --1.6 $\pm$  1.3\phn  &   0.07 \\
B2154+40  & 21:57:01.844  &   40:17:45.99  &  90.48 & --11.34 & 17.8 $\pm$ 0.8\phn    & 2.8 $\pm$ 1.0\phn     & --0.77 &   15.6  $\pm$  0.4\phn  &  --9.0 $\pm$  1.2\phn  &   0.09 \\
B2303+30  & 23:05:58.322  &   31:00:01.46  &  97.72 & --26.65 & 1.5 $\pm$ 2.3\phn     & --20.0 $\pm$ 2.2\phn  & --0.14 &   --7.5  $\pm$  2.1\phn  & --18.6 $\pm$  2.3\phn  & --0.11 \\
B2327--20 & 23:30:26.908  & --20:05:29.92  &  49.39 & --70.19 & 74.7 $\pm$ 1.9\phn    & 5.3 $\pm$ 2.9\phn     & 0.15   &   36.0  $\pm$  2.8\phn  & --65.7 $\pm$  1.9\phn  &   0.23 \\
\enddata
\tablenotetext{a}{Positions are J2000.0 coordinates at epoch 1999.5, and are accurate
to about 0.1~arcsecond.  Proper motion measurements are in milliarcseconds per year. No
corrections for differential galactic rotation have been made (see text).}
\end{deluxetable}

\begin{deluxetable}{lrrrrc}
\tablecaption{Comparisons with previous results\label{tab:vla_previous}}
\tablehead{
  \colhead{Pulsar}
& \colhead{$\mu_{\alpha}\,\cos\delta$}  
& \colhead{$\mu_{\delta}$}
& \colhead{$\Delta \mu_{\alpha}$} 
& \colhead{$\Delta \mu_{\delta}$}
& \colhead{reference}
\\
& \colhead{mas yr$^{-1}$}          
& \colhead{mas yr$^{-1}$}
& \colhead{$\sigma$}        
& \colhead{$\sigma$}
}
\startdata
B0149--16 &     3.1 $\pm$ 1.2 & --27.2 $\pm$ 2.0  &&&\\
          &     35 $\pm$ 47   &   --75 $\pm$ 19   &   0.67 & --2.51 & 1   \\
B0628--28 &   --44.6 $\pm$ 0.9  & 19.5 $\pm$ 2.2  &&&\\
          &   --37 $\pm$ 15   &      9 $\pm$ 7    &   0.50 & --1.43 & 1   \\
B0736--40 &   --14.0 $\pm$ 1.2  & 12.8 $\pm$ 2.1  &&&\\
          &   --57 $\pm$ 7    &   --21 $\pm$ 11   & --6.05 & --3.02 & 1   \\
B0919+06 &    18.8 $\pm$ 0.9    & 86.4 $\pm$ 0.7  &&&\\
         &   18.35 $\pm$ 0.06 &  86.56 $\pm$ 0.12 & --0.50 &   0.23 & 2  \\
         &      13 $\pm$ 29   &     64 $\pm$ 37   & --0.20 & --0.61 & 3    \\
B1237+25 &  --104.5 $\pm$ 1.1 & 49.4 $\pm$ 1.4    &&&\\
         & --106.82$\pm$0.17  & 49.92$\pm$0.18    & --2.08 & --0.37 & 4  \\
         &   --106 $\pm$ 4    &     42 $\pm$ 3    & --0.36 & --2.23 & 5    \\
B1508+55 &  --70.6 $\pm$ 1.6 & --68.8 $\pm$ 1.2   &&&\\
         &    --73 $\pm$ 4    &   --68 $\pm$ 3    & --0.55 &   0.24 & 5    \\
B1534+12 &   --7.6 $\pm$ 9.3  & --31.6 $\pm$ 10.1 &&&\\
         &   --1.5 $\pm$ 0.1  & --25.6 $\pm$ 1.5  &   0.65 &   0.59 & 6  \\
B1540--06 &  --17.4 $\pm$ 2.4  & --3.6 $\pm$ 2.7  &&&\\
          &   --11 $\pm$ 20   &    --1 $\pm$ 12   &   0.32 &   2.11 & 1   \\
B1541+09 &   --7.3 $\pm$ 1.0   & --4.0 $\pm$ 1.0  &&&\\
         &    --12 $\pm$ 4    &      3 $\pm$ 3    & --1.13 &   2.21 & 5    \\
B1642--03 &   --3.7 $\pm$ 1.5  & 30.0 $\pm$ 1.6   &&&\\
          &     41 $\pm$ 17   &   --25 $\pm$ 11   &   2.62 & --4.95 & 5    \\
B1718--02 &   --0.8 $\pm$ 3.7  & --25.7 $\pm$ 4.5 &&&\\
          &     26 $\pm$ 9    &   --13 $\pm$ 6    &   3.23 &   0.49 & 3    \\
B1946+35 &   --13.9 $\pm$ 0.8 & 0.7 $\pm$ 0.9     &&&\\
         &     --9 $\pm$ 7    &    --4 $\pm$ 8    &   0.88 & --0.58 & 3    \\
B2154+40 &   17.8 $\pm$ 0.8   & 2.8 $\pm$ 1.0 &&&\\
         &      18 $\pm$ 1    &    --3 $\pm$ 1    &   0.16 & --4.10 & 3    \\
B2303+30 &   1.5 $\pm$ 2.3   & --20.0 $\pm$ 2.2 &&&\\
         &      13 $\pm$ 8    &   --33 $\pm$ 6    &   1.38 & --2.03 & 5    \\
\enddata
\tablerefs{
(1) \citet{fgml97}
(2) \citet{ccl+01}
(3) \citet{hla93}
(4) \citet{bbgt02}
(5) \citet{las82}
(6) \citet{sac+98}
}
\end{deluxetable}

\begin{deluxetable}{lrrrrrr}
\tablewidth{0pt}
\tablecaption{Properties of six pulsars from Brisken et al. (2002) used in the analysis \label{tab:B2002}}
\tablehead{
    \colhead{Pulsar}
  & \colhead{$\alpha_{J2000}$}
  & \colhead{$\delta_{J2000}$}
  & \colhead{$\mu_{\alpha}$}
  & \colhead{$\mu_{\delta}$}
  & \colhead{$\pi$}
  & \colhead{$\tau_\mathrm{char}$}
\\
  &
  &
  & \colhead{(mas/yr)}
  & \colhead{(mas/yr)}
  & \colhead{(mas)}
  & \colhead{(Myr)}
}
\startdata
B0329+54 & $03^h32^m59^s.3862$ & $54^{\circ}34'43''.5051$ & $17.00\pm0.27$ & $-9.48\pm0.37$ & $0.94\pm0.11$ & 5.5 \\
B0950+08 & $09^h53^m09^s.3071$ & $07^{\circ}55'36''.1475$ & $-2.09\pm0.08$ & $29.46\pm0.07$ & $3.82\pm0.07$ & 17.5 \\
B1133+16 & $11^h36^m03^s.1829$ & $15^{\circ}51'09''.7257$ & $-73.95\pm0.38$ & $368.05\pm0.28$ & $2.80\pm0.16$ & 5.0 \\
B1929+10 & $19^h32^m13^s.9496$ & $10^{\circ}59'32''.4178$ & $94.82\pm0.26$ & $43.04\pm0.15$ & $3.02\pm0.09$ & 3.1 \\
B2020+28 & $20^h22^m37^s.0718$ & $28^{\circ}54'23''.0300$ & $-4.38\pm0.53$ & $-23.59\pm0.26$ & $0.37\pm0.12$ & 2.9 \\
B2021+51 & $20^h22^m49^s.8655$ & $51^{\circ}54'50''.3881$ & $-5.23\pm0.17$ & $11.54\pm0.28$ & $0.50\pm0.07$ & 2.7 \\
\enddata
\end{deluxetable}

\begin{deluxetable}{lcl}
\tablecaption{Braking index measurements to date\label{tab:brake}}
\tablehead{
  \colhead{Pulsar}
& \colhead{$n$}  
& \colhead{Reference}
}
\startdata
B0531+21 (Crab) & $2.51 \pm 0.01$ & Lyne et al. 1993 \\
B0540$-$69 & $2.04 \pm 0.02$ & Gouiffes et al. 1992 \\
B0833$-$45 (Vela) & $1.4 \pm 0.2$ & Lyne et al. 1996 \\
B1046$-$58 & $2.1 \pm 0.2$ & Urama 2002 \\
J1119$-$6127 & $2.91 \pm 0.05$ & Camilo et al. 2000 \\
B1509$-$58 & $2.837 \pm 0.001$ & Kaspi et al. 1994 \\
\enddata
\end{deluxetable}

\end{document}